\documentclass[conference]{IEEEtran}
\IEEEoverridecommandlockouts
\usepackage{cite}
\usepackage{amsmath,amssymb,amsfonts}
\usepackage{algorithmic}
\usepackage{graphicx}
\usepackage{textcomp}
\usepackage{xcolor}
\usepackage{caption}
\usepackage{float}
\usepackage{dblfloatfix}
\usepackage{subcaption}
\usepackage{array}
\usepackage{makecell}
\usepackage{multirow}
\usepackage{tikz}
\usepackage{pgfplots}
\usepackage{pgfplotstable}
\usepackage{subcaption}
\usepackage{soul}
\usepackage{color, colortbl}
\usepackage{url}
\usepackage{ragged2e}
\usepgfplotslibrary{statistics}
\usepackage{graphicx}
\usepackage{standalone}
\definecolor{yellow}{RGB}{175, 141, 195}

\DeclareUnicodeCharacter{2212}{\ensuremath{-}}
\def\BibTeX{{\rm B\kern-.05em{\sc i\kern-.025em b}\kern-.08em
    T\kern-.1667em\lower.7ex\hbox{E}\kern-.125emX}}

\newcommand{\mydiamond}{%
  \sbox0{$\lozenge$}%
  \usebox0\kern-.5\wd0\clap{\raisebox{.1ex}{\scalebox{.7}[1]{$-$}}}\kern.5\wd0%
}

\begin{document}

\title{NeuraLUT: Hiding Neural Network Density\\ in Boolean Synthesizable Functions\\
}

\author{\IEEEauthorblockN{Marta Andronic and George A. Constantinides}
\IEEEauthorblockA{Department of Electrical and Electronic Engineering \\
Imperial College London, UK\\
Email: \{marta.andronic18, g.constantinides\}@imperial.ac.uk}
}

\maketitle

\begin{abstract}
Field-Programmable Gate Array (FPGA) accelerators have proven successful in handling latency- and resource-critical deep neural network (DNN) inference tasks. Among the most computationally intensive operations in a neural network (NN) is the dot product between the feature and weight vectors. Thus, some previous FPGA acceleration works have proposed mapping neurons with quantized inputs and outputs directly to lookup tables (LUTs) for hardware implementation. In these works, the boundaries of the neurons coincide with the boundaries of the LUTs. We propose relaxing these boundaries and mapping entire sub-networks to a single LUT. As the sub-networks are absorbed within the LUT, the NN topology and precision within a partition do not affect the size of the lookup tables generated. Therefore, we utilize fully connected layers with floating-point precision inside each partition, which benefit from being universal function approximators, but with rigid sparsity and quantization enforced between partitions, where the NN topology becomes exposed to the circuit topology. Although cheap to implement, this approach can lead to very deep NNs, and so to tackle challenges like vanishing gradients, we also introduce skip connections inside the partitions. The resulting methodology can be seen as training DNNs with a specific FPGA hardware-inspired sparsity pattern that allows them to be mapped to much shallower circuit-level networks, thereby significantly improving latency. We validate our proposed method on a known latency-critical task, jet substructure tagging, and on the classical computer vision task, digit classification using MNIST. Our approach allows for greater function expressivity within the LUTs compared to existing work, leading to up to $4.3\times$ lower latency NNs for the same accuracy.
\end{abstract}

\begin{figure}
\centering
\begin{subfigure}{0.18\textwidth}
    \includegraphics[width=\textwidth]{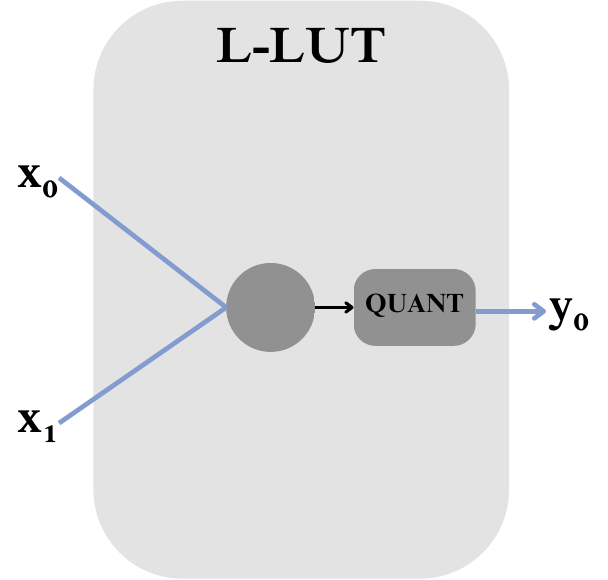}
    \caption{L-LUTs in LogicNets.}
    \label{fig:first}
\end{subfigure}
\hfill
\begin{subfigure}{0.22\textwidth}
    \includegraphics[width=\textwidth]{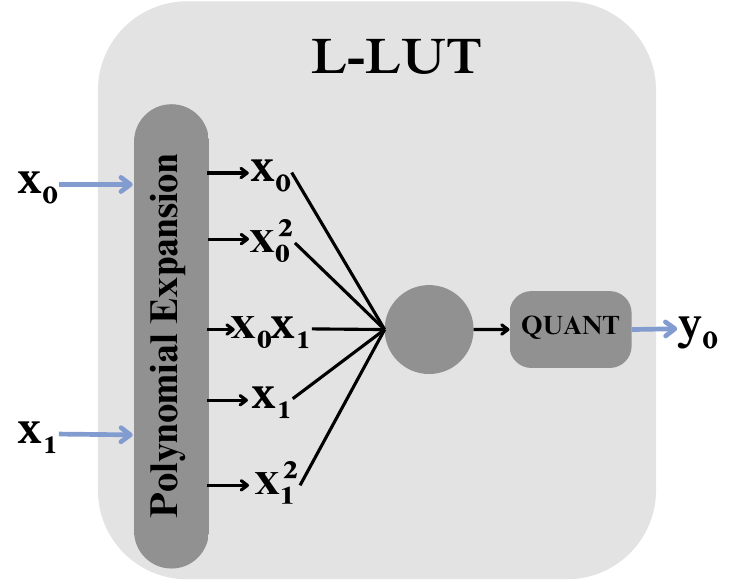}
    \caption{L-LUTs in PolyLUT.}
    \label{fig:second}
\end{subfigure}
\hfill
\begin{subfigure}{0.22\textwidth}
    \includegraphics[width=\textwidth]{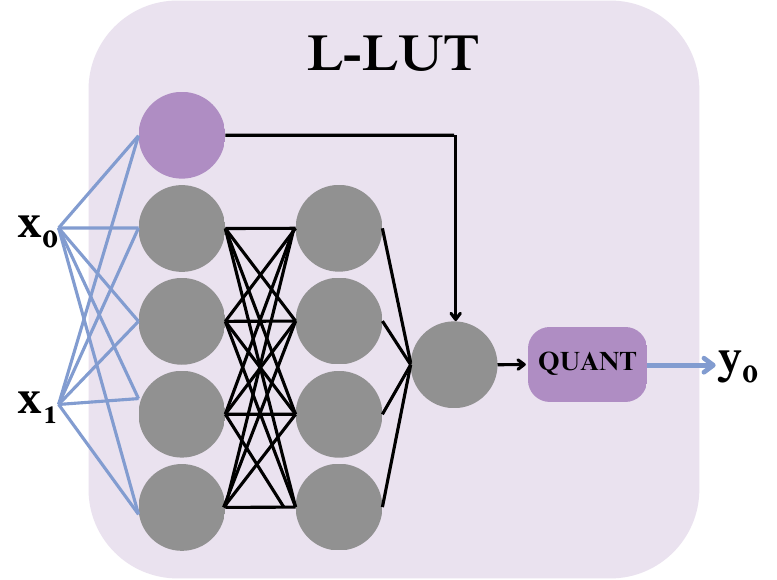}
    \caption{L-LUTs in NeuraLUT.}
    \label{fig:third}
\end{subfigure}
        
\caption{Gray circle: Affine transformation + ReLU. Purple circle: Affine transformation for residual connections. Blue lines: Low precision. Black lines: Full precision.}
\label{fig:figures}
\end{figure}

\section{Introduction and Motivation}
The deployment of DNNs on the edge has led to many breakthroughs across a wide range of domains, for example, in particle collision~\cite{duarte}, cybersecurity~\cite{murovic} and X-ray classification~\cite{cookiebox}. Edge devices, characterized by limited resources, demand specialized solutions that can deliver efficient and real-time inference without compromising accuracy. However, designing deep learning models capable of meeting the stringent requirements of edge devices has proven to be a great challenge. This difficulty arises from the inherent high computational complexity and substantial footprint of these models. In response to these challenges, research efforts come from both hardware and software perspectives~\cite{edge}.

Custom hardware accelerators have proven to reach performance levels that were previously unreachable by general-purpose processors. The efficacy of these hardware-efficient accelerators stems from innovative approximation methods employed to preserve accuracy on smaller, lower-precision, and sparse models~\cite{esurvey}. Examples of such techniques include parameter pruning, network
quantization, and knowledge distillation. Moreover, FPGAs are ideal for prototyping and deploying cutting-edge DNNs because their reconfigurability allows for rapid design iteration~\cite{esurvey}.

LUT-based NNs have emerged as an alternative to binary neural networks (BNNs) due to the limitations of the latter in fully leveraging FPGA resources. For example, BNNs do not efficiently utilize $K$-input LUTs~\cite{lutnet1}. Prior LUT-based NNs include PolyLUT~\cite{poly}, LUTNet~\cite{lutnet1}, LogicNets~\cite{logicnets} and NullaNet~\cite{nullanet}. The main motivation of these works was to utilize the fact that LUTs are able to implement $K$-input Boolean operations and encapsulate inside a single LUT more complex functions to increase the efficiency of traditional NNs.

LUTNet~\cite{lutnet1} is a NN architecture that replaces the XNORs in BNNs with trainable $K$-input LUTs. We refer to these LUTs as Physical-LUTs (P-LUTs) to emphasize their direct correspondence to native FPGA $K$-input LUTs. LUTNet is not limited to training weighted sums and is able to train sums of arbitrary Boolean functions of $K$ activations. However, LUTNet maintains the exposed datapaths of a BNN, including the popcount operations, rather than also packing this functionality into the trained LUTs, and it inherits BNNs' restriction of having single-bit activations. 

Following~\cite{poly}, we refer to LUTs of arbitrary size as Logical-LUTs (L-LUTs) to underline the fact that they can exceed the number of P-LUT inputs, in which case they get implemented by the synthesis tools as circuits of multiple P-LUTs. PolyLUT~\cite{poly}, LogicNets~\cite{logicnets}, and NullaNet~\cite{nullanet} absorb the full computation of a neuron inside a single L-LUT, leaving no exposed datapaths except the connections between layers, and allow multi-bit precision. Therefore, the NN gets converted to a network of L-LUTs. What differentiates these works is what functions get encapsulated inside the L-LUTs: LogicNets and NullaNet encapsulate traditional linear + activation neurons (Figure~\ref{fig:first}), while PolyLUT encapsulates multivariate polynomials + activation neurons (Figure~\ref{fig:second}).

\begin{figure}[tb]
\centerline{\includegraphics[width=0.8\columnwidth]{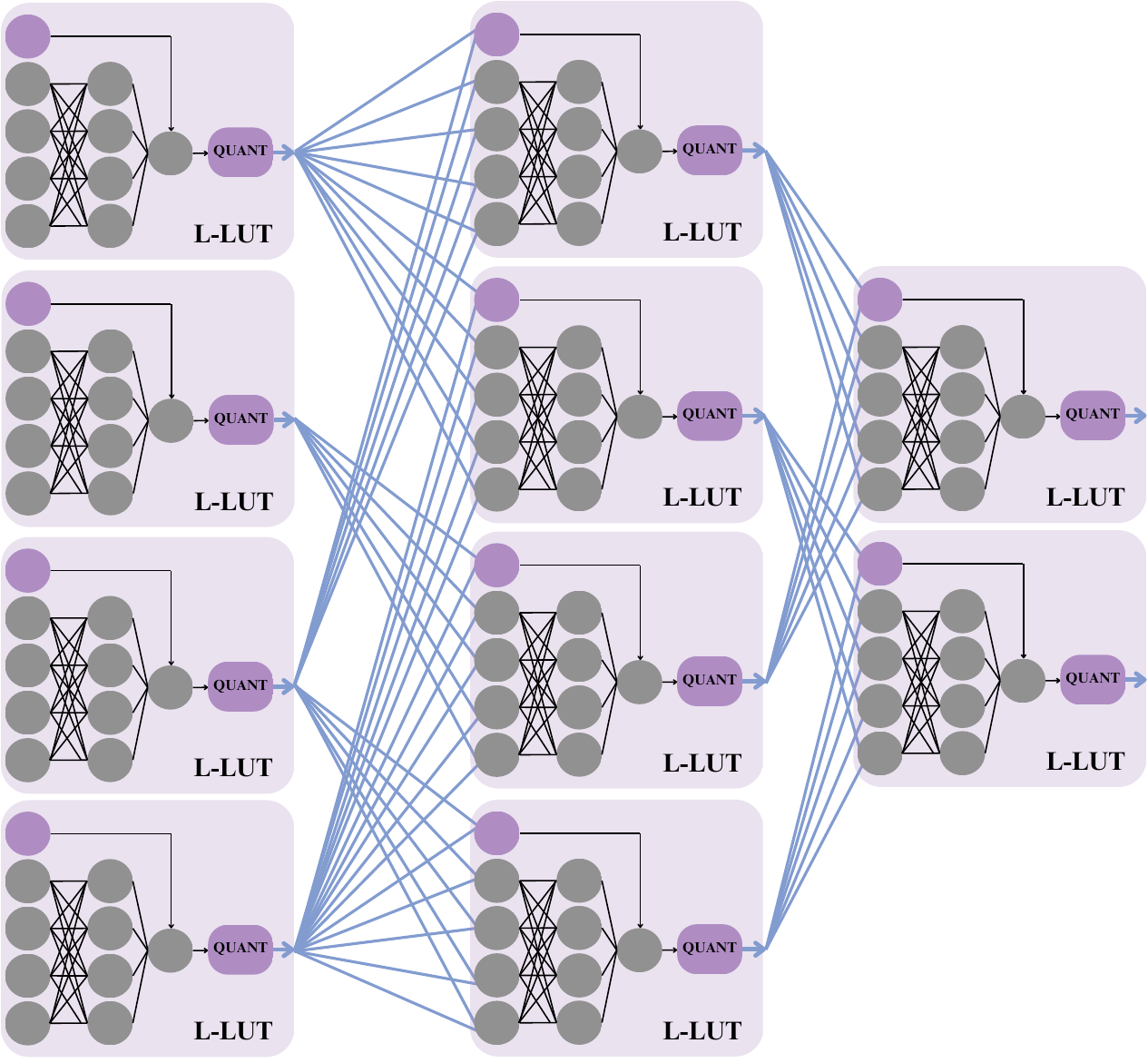}}
\caption{High-level view of NeuraLUT's architecture. Sparsely connected dense sub-networks with skip-connections.}
\label{fig:flow}
\end{figure}

In latency-critical applications two considerations have to be regarded: the first one is reducing the latency associated with each layer, and the second one is reducing the number of layers. PolyLUT~\cite{poly} and LogicNets~\cite{logicnets} have succeeded in reducing the number of clock cycles associated with each layer to just one while maintaining a high frequency. However, the accuracy achievable degrades considerably as networks become very shallow. This develops into a significant limitation for the traditional success of DNNs, where a higher number of layers is often correlated with improved performance. Therefore, we propose designing deep NNs with specific sparsity patterns that resemble sparsely connected dense partitions, enabling the encapsulation of sub-networks entirely within a single L-LUT (Figure~\ref{fig:third}). The advantage is that the network can reach greater function expressibility while keeping the number of circuit-level layers minimal (Figure~\ref{fig:flow}), and by hiding these sub-networks inside LUTs only the quantization of inputs and outputs is required, while the rest of the parameters maintain full precision. However, substantially increasing the depth of each partition, even though it does not impact the implementation complexity, does impact the training complexity, and can lead to vanishing/exploding gradients, an issue for which we also provide a solution.

In this paper, we present the following novel contributions:
\begin{itemize}
    \item{We introduce NeuraLUT, an open-source\footnote{\url{https://github.com/MartaAndronic/NeuraLUT}} framework designed to leverage the underlying structure of the FPGA architecture by hiding dense and full precision sub-networks within synthesizable Boolean lookup tables.
    }
    \item{We demonstrate that utilizing sub-networks in the places where prior works have used linear or polynomial functions enhances the representational capacity of NNs, facilitating significant reductions in the depth and width of the circuit-level model architecture.}
    \item{We enhance the training by integrating skip-connections in our sub-networks which facilitate the flow of gradients, promoting stable and efficient learning without affecting the circuit topology, as these connections are also hidden inside the LUTs.}
    \item{We assess NeuraLUT using two datasets with distinct applications. Our results demonstrate that, for comparable accuracies, NeuraLUT achieves the lowest latencies, with reductions on MNIST of $1.3\times$ against PolyLUT~\cite{poly}, and on the jet tagging dataset of $1.6\times$ against PolyLUT and $4.3\times$ against LogicNets~\cite{logicnets}.}
\end{itemize}

\section{Background}
Designing NNs that run in real-time and have minimal area footprint while maintaining good accuracy requires rethinking the model design to optimize performance on the target hardware. Previous efforts have centered on co-designing NN architectures for dedicated hardware platforms, which involves an intrinsic development loop between model architecture design, training, and deployment. Prior co-design works that have been specialized for FPGA platforms can be split into three categories based on their main computational block: DSP-based, XNOR-based, and LUT-based.

\subsection{DSP-based Neural Networks}
\texttt{hls4ml}~\cite{duarte} is an open-source framework that specializes in mapping NNs to FPGAs for low-latency applications. Duarte \textit{et al.}~\cite{duarte} leverage the \texttt{hls4ml} framework to generate latency-efficient fully-unrolled and rolled designs. However, the work utilizes high network precision, which leads to significant Digital Signal Processing (DSP) utilization. Fahim \textit{et al.}~\cite{fahim} utilize \texttt{hls4ml} while incorporating techniques from prior works, such as boosted decision trees~\cite{summers} and quantization-aware training~\cite{coelho}. They further propose quantization-aware pruning to achieve higher performance and power efficiency.
\subsection{XNOR-based Neural Networks}
FINN~\cite{finn} is an open-source framework, originally designed for building high-performing BNN accelerators on FPGAs. FINN introduces hardware-specific optimizations such as replacing additions with popcount operators, replacing batch normalization and activation with thresholding, and replacing max-pooling with Boolean ORs. Another work that focuses on mapping binary or ternary NNs is proposed by Ngadiuba \textit{et al.}~\cite{hls4ml} and it utilizes \texttt{hls4ml}.
\subsection{LUT-based Neural Networks}
LUTNet~\cite{lutnet2} innovatively departed from traditional machine learning operators by replacing the XNORs of a BNN with learned Boolean functions. At inference, each Boolean function is efficiently computed using a single LUT on the FPGA, thus taking advantage of the resources available on the FPGA fabric. While the XNORs are operating on a single input feature, the LUTs operate on $K$ features. Therefore, the structure of LUTNet supports multiple occurrences of each input in the neuron function allowing redundant operations to be removed through network pruning while maintaining high accuracy. However, LUTNet displays exponential scaling of training parameters with the size of LUT inputs, which proves feasible only in the context of binary activations.

NullaNet~\cite{nullanet} and LogicNets~\cite{logicnets} present neurons as multi-input multi-output Boolean functions. In NullaNet, the functions undergo Boolean logic minimization and, in order to manage computational resources, output values are determined selectively for specific input combinations, leaving the remaining outputs as don't-care conditions. LogicNets, on the other hand, employs high sparsity to address the drawback of NullaNet's lossy truth table conversion. To achieve this high sparsity, LogicNets applies an \textit{a priori} random sparsity technique, which is supported by expander graph theory~\cite{expander}. This design approach allows the reduction of the input vector size of each neuron to a user-defined fan-in $F$ and combats the exponential growth of truth tables, thus leading to efficient implementations. Therefore, the number of trainable parameters associated with each neuron has complexity $\mathcal{O}(F)$ (Table~\ref{table:params}) in LogicNets.

PolyLUT~\cite{poly} is also a NN architecture which absorbs all the operations performed by a neuron within a L-LUT, but it distinctively expands the feature vector at each neuron by incorporating all monomials up to a user-defined degree $D$. Consequently, the number of trainable parameters has complexity $\mathcal{O}\left(\left(F+D \atop D \right)\right)$ (Table~\ref{table:params}). This expansion allows the model to capture complex relationships within the data through higher-degree polynomial expressions. Notably, PolyLUT integrates multiplicative interactions within a LUT-based model, but avoids the need for additional multiplication hardware by encapsulating everything inside the L-LUT. In contrast to LogicNets, which computes a continuous piecewise linear function, PolyLUT computes a continuous piecewise polynomial function. The increased function complexity within each layer further contributes to the reduction of required layers for achieving a given accuracy.

\subsection{Network in Network}
The concept of micro-networks within a larger network has been proposed before, in a very different context: the design of CNN structures and as part of the sliding window operation. For example, Lin \textit{et al.}~\cite{NIN} introduced Network in Network (NIN). In the Network in Network~\cite{NIN} architecture, multilayer perceptrons (MLPs) are utilized as a replacement for the traditional linear filters in convolutional layers. 

The use of MLPs in~\cite{NIN} is motivated by the desire for a universal function approximator that can capture more abstract representations of latent concepts without relying on assumptions about their distributions. Moreover, MLPs are chosen for their compatibility with the structure of CNNs. They can be seamlessly integrated into the network and trained using back-propagation, facilitating the end-to-end learning process. Finally, MLPs address limitations associated with traditional convolutional layers in capturing abstract representations of latent concepts.

In contrast to NIN, NeuraLUT utilizes MLPs to capture more meaningful relationships within a L-LUT. This approach capitalizes on the benefits of MLPs as universal function approximators and their compatibility with traditional NN training techniques.

\section{Methodology}
\label{section:methodology}

\begin{table*}[t]
\caption{Breakdown of the main L-LUT characteristics. $F$ is the L-LUT fan-in, $D$ is the degree of the polynomials, $L$ is the depth of the sub-networks, and $N$ is the width of the hidden layers of the sub-networks.}
\begin{center}
\renewcommand{\arraystretch}{1.5} 
\setlength{\tabcolsep}{12pt}
\begin{tabular}{cccc}
\hline
&\multirow{2}{*}{\textbf{Function hidden inside each L-LUT}}&\multirow{2}{*}{\textbf{No. of parameters}}&\multirow{2}{*}{\textbf{Scaling type}}\\&&&\\
\hline
\hline
\multirow{2}{*}{\textbf{LogicNets}~\cite{logicnets}}&\multirow{2}{*}{Linear + Activation}&\multirow{2}{*}{$\mathcal{O}(F)$}&\multirow{2}{*}{Linear in $F$}\\&&&\\
\hline
\multirow{2}{*}{\textbf{PolyLUT}~\cite{poly}}&\multirow{2}{*}{Multivariate polynomial + Activation}&\multirow{2}{*}{$\mathcal{O}\left(\left(F+D \atop D \right)\right)$}&Polynomial in $F$ \\&&&(for fixed $D$)\\

\hline
\multirow{2}{*}{\textbf{NeuraLUT} (this work)}&\multirow{2}{*}{Arbitrary neural network}
&\multirow{2}{*}{$\mathcal{O}(LN^2+(F+L)N)$}&Linear in $F$
\\&&&(for fixed $N$, $L$)\\
\hline
\end{tabular}
\label{table:params}
\end{center}
\end{table*}

The key novelty of our work lies in the design of LUT-based NNs, where each LUT is capable of performing functions more powerful than traditional linear mappings. The most important observation is that LUTs have the capability to implement arbitrary functions, and they have been used in machine learning to implement linear transformations as shown in LogicNets or polynomials as demonstrated in PolyLUT.

While linear functions are simpler, it has been shown in PolyLUT that they underutilize the full potential of LUTs, resulting in less efficiency compared to more complex functions like polynomials. However, multivariate polynomial functions come with potentially exponentially increasing degrees of freedom, and as observed in PolyLUT, NNs of this kind are challenging to train, with diminishing returns when the degree exceeds two~\cite{poly}.

Yet, there exists another universal function approximator with the advantage of ease of training without changing existing training frameworks: the multilayer perceptron~\cite{cybenko, hornik}. Consequently, we embed MLPs within LUT functions, increasing the function expressivity of each L-LUT, while keeping the number of L-LUTs fixed.

\subsection{Tackling LUT size}
NeuraLUT manages the fact that the size of a LUT is exponential in its number of inputs by containing regions of high NN density inside the L-LUTs while keeping the circuit-level model (between L-LUTs) highly sparse. Moreover, NeuraLUT adopts from LogicNets the \textit{a priori} sparsity random technique which restricts the number of inputs to each L-LUT to a fan-in parameter $F$ and restricts the bit-width of the circuit-level inputs to a bit-width parameter $\beta$.

\subsection{Skip-connections}
Training particularly deep NNs presents challenges due to the vanishing gradient problem~\cite{bengio}~\cite{glorot}. This phenomenon arises when gradients diminish significantly or ``vanish" as they propagate backward through the network during training. The circuit-level model NNs for ultra-low latency applications have limited depth and the problem of vanishing gradients has not been deemed as an issue in prior works. However, in the NeuraLUT context, the hidden sub-networks have high depth relative to the number of L-LUT inputs and are hard to train unless residual connections are employed. Residual connections mitigate this problem by accumulating the outputs of some layers with the activations from previous layers~\cite{resnet}. The advantage of using residual connections within the network partitions is that they come at a minimal cost because they can also be encapsulated inside the L-LUT.

\subsection{Function hidden inside the L-LUT}
NeuraLUT's performance boost comes from the representational power of the hidden dense residual NN as a function of depth, width and residual connection step. To describe the expressive power of our LUTs precisely, we require some notation. We denote a NN hidden in an L-LUT by $\mathcal{N}$, which is characterized by the following integers: $L$ representing the depth of $\mathcal{N}$, $n_{\text{in}}=n_0$ the input size, $n_1,n_2,...,n_L=n_{\text{out}}$ the widths of the layers, and $S$ quantifying the number of layers that are skipped by the residual connections. $S=0$ is a special case for no skip connections. Given an activation function $\phi$, we say that $\mathcal{N}$ computes the following function prior to the quantized activation (where, for simplicity, we assume that $L$ is a multiple of $S\neq0$, and where $\circ$ denotes function composition):
\begin{equation}
    f_{\mathcal{N}} = F_{\frac{L}{S}} \circ \phi \circ F_{\frac{L}{S}-1} \circ \cdots \circ F_{2} \circ \phi \circ F_{1},
\end{equation}
where $F_i:\mathbb{R}^{n_{S(i-1)}} \rightarrow \mathbb{R}^{n_{Si}}$. This expresses the function computed via composing chunks of layers, each of which is equipped with a skip connection $R_i$, as below:

\begin{equation}
    F_i(\textbf{x}) = \Hat{F}_i(\textbf{x}) + R_i(\textbf{x}),
\end{equation}
where $\Hat{F}_i(\textbf{x}):\mathbb{R}^{n_{S(i-1)}} \rightarrow \mathbb{R}^{n_{Si}}$ is a multi-layer perceptron, such that

\begin{equation}
    \Hat{F}_i(\textbf{x}) = A_{Si} \circ \phi \circ A_{Si-1} \circ \cdots \circ \phi \circ A_{Si-S+1},
\end{equation}
and where $R_i:\mathbb{R}^{n_{S(i-1)}} \rightarrow \mathbb{R}^{n_{Si}}$ and $A_i:\mathbb{R}^{n_{i-1}} \rightarrow \mathbb{R}^{n_{i}}$ are affine transformations. In this work, we focus on
\begin{equation}
\begin{split}
    \phi(\textbf{x}) &= \text{ReLU}(x_1,\dotsc,x_k)\\&= (\max(0,x_1),\dotsc,\max(0,x_L)).
\end{split}
\end{equation}
Each L-LUT has $n_{in}=F$ and $n_{out}=1$. Moreover, in NeuraLUT all the widths of the hidden layer are equal for simplicity, \textit{i.e.} $n_{2}=\cdots=n_{L-1}=N$. Therefore, the number of trainable parameters of $\mathcal{N}$ will be equal to the number of weights and bias terms for each layer and residual connection. We denote the total number of trainable parameters of $\mathcal{N}$ with $T_\mathcal{N}$ and the total number of trainable parameters associated with all $A_i$ and $R_i$ with $T_A$ and $T_R$, respectively. Additionally, we define $T$ to be the function that returns the number of trainable parameters associated with an affine transformation $X:\mathbb{R}^{d_1} \rightarrow \mathbb{R}^{d_2}$, \textit{i.e.} $T(X) = d_1 \cdot d_2+d_2$.
\begin{equation}
\begin{split}
    T_A &= T(A_1) + T(A_2)+\cdots+ T(A_L)\\
    &=\begin{cases} (F \cdot 1 + 1), & \text{if } L=1 \\ (F \cdot N + N) + (N \cdot 1 + 1), & \text{if } L=2 \\ \begin{split}(F \cdot N + N) + (N \cdot 1 + 1) \\+ (N\cdot N + N)(L-2), \end{split}& \text{if } L>2.\end{cases}\\
    &=\begin{cases} F + 1, & \text{if } L=1 \\ (F+2)N + 1, & \text{if } L=2 \\ (L-2)N^2 + (F+L)N + 1, & \text{if } L>2.\end{cases}
\end{split}
\end{equation}
Similarly,
\begin{equation}
\begin{split}
    T_R &= T(R_1) + T(R_2)+\cdots+ T(R_\frac{L}{S})\\
    &=\begin{cases} F + 1, & \text{if } \frac{L}{S}=1 \\ (F+2)N + 1, & \text{if } \frac{L}{S}=2 \\ (\frac{L}{S}-2)N^2 + (F+\frac{L}{S})N + 1, & \text{if } \frac{L}{S}>2\end{cases}.
\end{split}
\end{equation}
Therefore, as seen in Table~\ref{table:params}, the total number of trainable parameters for an $F-$input L-LUT is
\begin{equation}
\begin{split}
T_{\mathcal{N}} &= T_A + T_R\\&= \mathcal{O}(LN^2+(F+L)N).
\end{split}
\end{equation}
This analysis reveals two main scalability advantages over PolyLUT, as apparent from the table. Firstly, the scaling in the fan-in for NeuraLUT is linear for fixed expressibility parameters $N$, $L$, whereas for PolyLUT it is polynomial for fixed expressibility parameter $D$. Secondly, the scaling in the expressibility parameters themselves is polynomial, whereas for PolyLUT it is exponential due to the combinatorial expression. Additionally, when $N$=$L$=$1$, and $S$=$0$, NeuraLUT is equivalent to LogicNets, making it, like PolyLUT, a strict generalization of LogicNets.
\begin{figure}[t]
\centerline{\includegraphics[width=1.05\columnwidth]{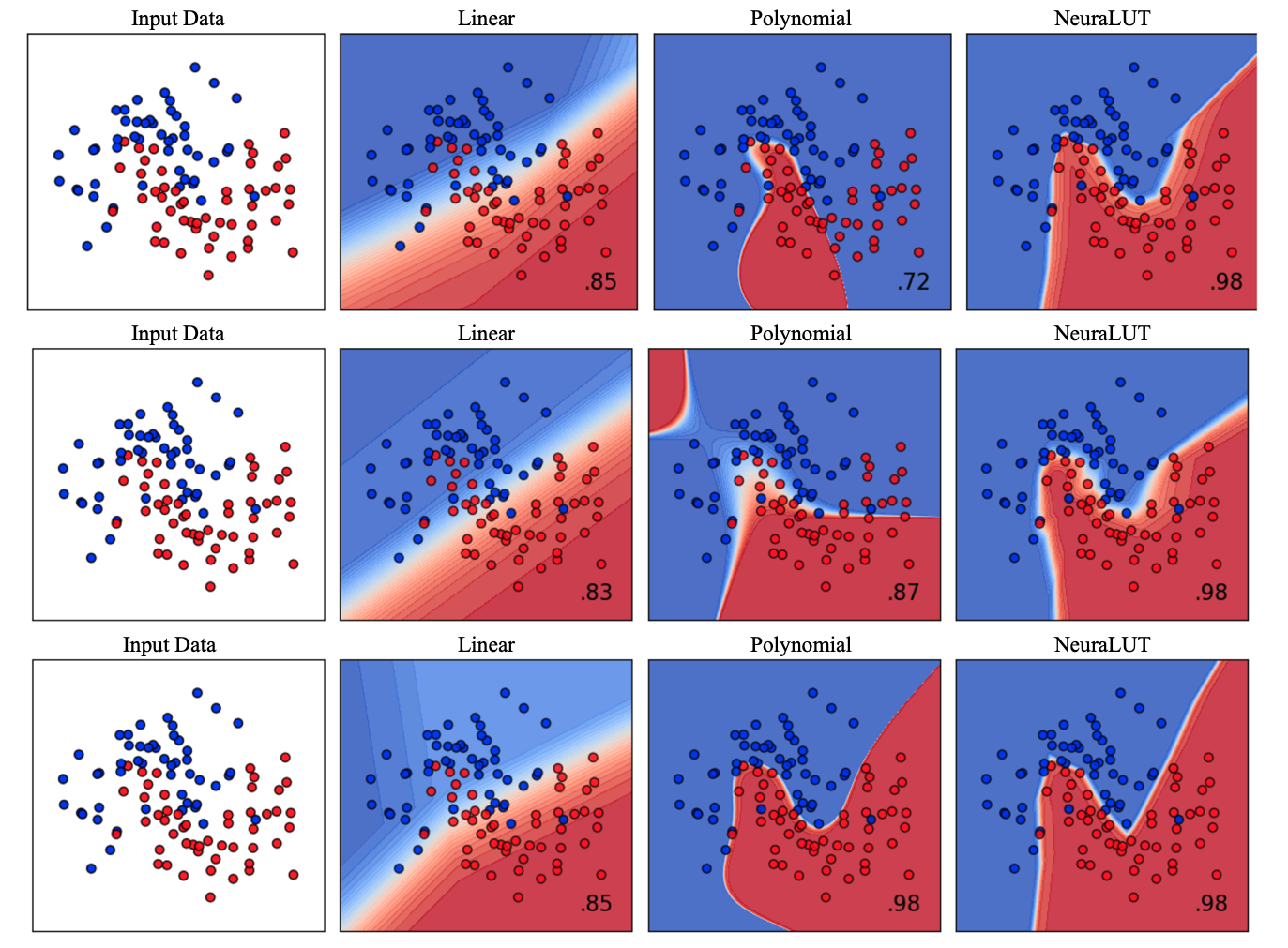}}
\caption{Visualization of decision boundaries. Classifier comparison across three different seeds.}
\label{fig:toy}
\end{figure}
\begin{figure*}[t]
\centerline{\includegraphics[width=2.05\columnwidth]{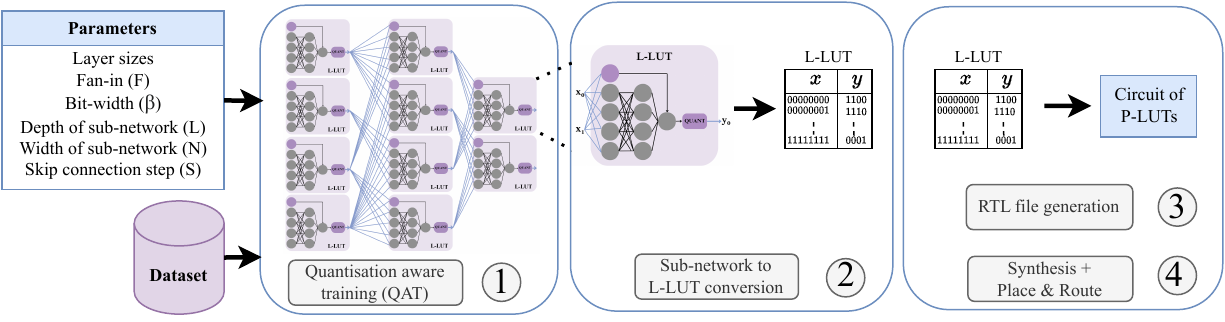}}
\caption{Visualization of NeuraLUT's toolflow, consisting of four stages.}
\label{fig:toolflow}
\end{figure*}

\subsection{Expressive power}

In NeuraLUT, the L-LUT incorporates a linear NN that independently computes a continuous piecewise linear function. Thus, the network also computes a continuous piecewise linear function, but with more (potentially many more) piecewise regions compared to LogicNets when $L>1$, because the number of regions can be exponential in $L$.

To visually illustrate the benefits of NeuraLUT's methodology, we train a $3$-layer toy NN in three configurations. In the first configuration, each neuron contains a single linear function capturing the expressive power of LogicNets~\cite{logicnets} and in the second configuration a single polynomial capturing PolyLUT. The third configuration illustrates our work, NeuraLUT, when replacing each neuron with a $2$-layer NN ($L=2$, $S=0$). We trained these models with various seeds on a toy dataset featuring two semicircles, as shown in Figure~\ref{fig:toy}. The contour graphs showcase the classification boundaries, where the blue and red regions indicate different output classifications of the network input and the white area serves as the decision boundary.

The visualization illustrates that compared to LogicNets, NeuraLUT has superior capability in discerning intricate data distributions with high accuracy when using a highly restricted number of layers. Compared to PolyLUT, we observed a trend across multiple seed runs: NeuraLUT consistently converges to highly accurate solutions, while the polynomial network may yield impressive classifications like the one at the bottom of Figure~\ref{fig:toy}, it may reach an inferior solution, as illustrated in the middle, or it can even fall short of the linear case's accuracy, as evident at the top. Also compared to PolyLUT, NeuraLUT’s training is simpler, more stable, and relies only on standard multi-layer perceptrons with skip-connections, well supported by existing training frameworks like PyTorch.

\subsection{Toolflow}
NeuraLUT extends LogicNets' toolflow~\cite{logicnets}, facilitating the DNN training, conversion to L-LUTs, RTL file generation, and hardware compilation. Modifications were made to the training implementation to accommodate the structure of NeuraLUT. The high-level view of the toolflow is illustrated in Figure~\ref{fig:toolflow}.

\subsubsection{Quantization-aware training (QAT)}
The DNN training is carried out using PyTorch. The initial step in the pipeline involves specifying learning-specific parameters, such as the learning rate, as well as topology parameters, as illustrated in Figure~\ref{fig:toolflow}. The hyperparameters $L$, $N$, and $S$ introduced in this framework are constant across all sub-networks and describe their topology within the L-LUTs. The layer sizes, fan-in and bit-width refer to the circuit-level topology.

Once these parameters are defined and the dataset is selected, the model is trained employing Decoupled Weight Decay Regularization~\cite{reg} and Stochastic Gradient Descent with Warm Restarts~\cite{sgd}. Additionally, the inputs and outputs of each sub-network undergo batch normalization and quantization using Brevitas~\cite{brevitas} quantized activation functions, which incorporate learned scaling factors. Each model undergoes training for $1000$ epochs for the jet substructure tagging dataset and $500$ epochs for the MNIST dataset.

\subsubsection{Sub-network to L-LUT conversion}
The second stage in the pipeline involves converting each sub-network into an L-LUT. This process is carried out in PyTorch by first generating all input combinations based on their specified bit-width and then evaluating the sub-network function on each of these combinations through inference. The number of entries in the L-LUT is $2^{\beta F}$, as in LogicNets, with only the content of the lookup table differing.

\subsubsection{RTL file generation}
As part of the same PyTorch framework, the network is automatically converted to Verilog RTL and each L-LUT is written out as read-only memories (ROMs) with registers at the output.

\subsubsection{Synthesis and Place \& Route}
To compile the Verilog RTL files, we utilize Vivado $2020.1$, selecting the \texttt{xcvu9p-flgb2104-2-i} FPGA part, to enable direct comparison to both LogicNets and PolyLUT. To ensure consistency, we compile the projects using Vivado's  \texttt{Flow\_PerfOptimized\_high} settings and execute synthesis in the \texttt{Out-of-Context} mode. The target clock periods are set at $1$ or $2$ ns depending on the network size.

\section{Experimental Results}
To evaluate our network architecture, we train our models on an ultra-low latency and size-critical dataset, the jet substructure tagging dataset as presented in~\cite{duarte}, and on the MNIST dataset~\cite{mnist}.

The jet substructure tagging dataset originates from CERN, and low latency is vital at their Large Hadron Collider (LHC). The deployment of ultra-low latency machine learning models at the LHC enhances the experiments to preserve potential new physics signatures that would otherwise be lost as part of the initial stage of triggering. Therefore, we showcase NeuraLUT's effectiveness on this dataset which contains $16$ substructure properties to classify $5$ types of jets, offering potential applications in high-energy physics. To evaluate our method, we also utilize the MNIST dataset, featuring $28\times28$ pixel images of handwritten digits flattened into $784$-dimensional inputs, with $10$ output classes representing each digit.

Initially, we perform a case study on the MNIST dataset, aiming to understand NeuraLUT's training outcomes, including test accuracy, latency, and area footprint. Our primary goal is to showcase how integrating fully connected sub-networks with skip-connections inside L-LUTs can increase training effectiveness and enable the generation of ultra-low latency and minimal area FPGA implementations. Focusing on the MNIST dataset provides valuable insights into the advantages of our approach, after which we evaluate multiple datasets and compare them against the state of the art to highlight the effectiveness of our method on different tasks.

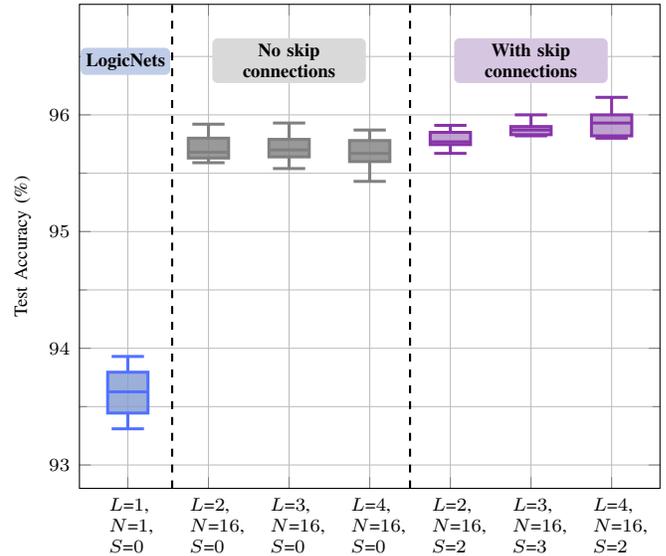
\begin{figure}[t]
\hspace{-0.03\columnwidth}
    \centering
    \begin{tikzpicture}
  \definecolor{yellow}{RGB}{175, 141, 195}
  \definecolor{dyellow}{RGB}{136, 52, 171}
  \definecolor{blue}{RGB}{130, 156, 208}
  \definecolor{red}{RGB}{255, 87, 87}
  \definecolor{dblue}{RGB}{82, 113, 255}
  \tikzstyle{every node}=[font=\scriptsize]
  \begin{axis}[
      width=1.05\columnwidth,
      ylabel={Test Accuracy (\%)},
      grid=both,
      boxplot/draw direction=y,
      xtick={1,2,3,4,5,6,7},
      yticklabel style={rotate=0},
      xticklabel style={rotate=0, text width=5mm, font=\scriptsize,},
      ylabel near ticks,
      yminorticks=true,
      minor y tick num=1,
      ymin= 92.8,
      xmax = 7.6,
      xmin = 0.4,
      boxplot/box extend=0.5,
      xticklabels={{$L$=$1$, $N$=$1$, $S$=$0$}, {$L$=$2$, $N$=$16$, $S$=$0$}, {$L$=$3$, $N$=$16$, $S$=$0$}, {$L$=$4$, $N$=$16$, $S$=$0$}, {$L$=$2$, $N$=$16$, $S$=$2$}, {$L$=$3$, $N$=$16$, $S$=$3$}, {$L$=$4$, $N$=$16$, $S$=$2$}},
      cycle list={{blue},{blue}},
    ]
    \addplot+[mark = *,fill=blue, draw=dblue, fill opacity=0.8,line width=0.4mm,
      boxplot prepared={lower whisker=93.31, lower quartile=93.445, median=93.627, upper quartile=93.795, upper whisker=93.93},
    ] coordinates{};
    \addplot+[mark = *,fill=gray, draw=gray, fill opacity=0.8,line width=0.4mm,
      boxplot prepared={lower whisker=95.59, lower quartile=95.63, median=95.68, upper quartile=95.8, upper whisker=95.92},
    ] coordinates{};
    \addplot+[mark = *,fill=gray, draw=gray, fill opacity=0.8,line width=0.4mm,
      boxplot prepared={lower whisker=95.54, lower quartile=95.64, median=95.7, upper quartile=95.79, upper whisker=95.93},
    ] coordinates{};
    \addplot+[mark = *,fill=gray, draw=gray, fill opacity=0.8,line width=0.4mm,
      boxplot prepared={lower whisker=95.43, lower quartile=95.6, median=95.67, upper quartile=95.78, upper whisker=95.87},
    ] coordinates{};
    \addplot+[mark = *,fill=yellow, draw=dyellow, fill opacity=0.8,line width=0.4mm,
      boxplot prepared={lower whisker=95.67, lower quartile=95.745, median=95.769, upper quartile=95.85, upper whisker=95.91},
    ] coordinates{};
    \addplot+[mark = *,fill=yellow, draw=dyellow, fill opacity=0.8,line width=0.4mm,
      boxplot prepared={lower whisker=95.82, lower quartile=95.83, median=95.87, upper quartile=95.90, upper whisker=96},
    ] coordinates{};
    \addplot+[mark = *,fill=yellow, draw=dyellow, fill opacity=0.8,line width=0.4mm,
      boxplot prepared={lower whisker=95.8, lower quartile=95.82, median=95.93, upper quartile=96, upper whisker=96.15}, 
    ] coordinates{};
    \draw[black, dashed, thick] (axis cs:4.5,\pgfkeysvalueof{/pgfplots/ymin}) -- (axis cs:4.5,\pgfkeysvalueof{/pgfplots/ymax});
    \draw[black, dashed, thick] (axis cs:1.55,\pgfkeysvalueof{/pgfplots/ymin}) -- (axis cs:1.55,\pgfkeysvalueof{/pgfplots/ymax});
    \addplot [gray,-]
    coordinates {(0.97, 96.57)}
    node[text=black, font=\tiny, above, yshift=-0.36cm] at (12,28.6) [pos=0.45,font=\scriptsize, text width=10.5mm, fill=blue!50!white, rounded corners=2pt, inner sep=2pt,  align=center]{\textbf{LogicNets}};;
    \addplot [gray,-]
    coordinates {(3, 96.5)}
    node[text=black, font=\tiny, above, yshift=-0.36cm] at (12,28.6) [pos=0.45,font=\scriptsize, text width=19mm, fill=gray!30!white, rounded corners=2pt, inner sep=2pt,  align=center]{\textbf{No skip connections}};;
    \addplot [gray,-]
    coordinates {(6, 96.5)}
    node[text=black, font=\tiny, above, yshift=-0.36cm] at (12,28.6) [pos=0.45,font=\scriptsize, text width=19mm, fill=yellow!50!white, rounded corners=2pt, inner sep=2pt,  align=center]{\textbf{With skip connections}};;
  \end{axis}
\end{tikzpicture}
    \caption{Ablation study on MNIST across 10 seeds. Blue: baseline, Gray: NeuraLUT without skip-connections, Purple: standard NeuraLUT.  All models have a fixed circuit-level architecture with ($256,100,100,100,100,10$) L-LUTs.}
    \label{fig:box_plot}
\end{figure}

\begin{figure}[tbp]
	\centering
 \hspace{-1.1cm}
        \resizebox{1.1\columnwidth}{!}{\definecolor{yellow}{RGB}{175, 141, 195}
\definecolor{dyellow}{RGB}{136, 52, 171}
\definecolor{blue}{RGB}{130, 156, 208}
\definecolor{red}{RGB}{255, 87, 87}
\definecolor{dblue}{RGB}{82, 113, 255}

\begin{tikzpicture}
  \pgfplotsset{compat=1.5}
  \tikzstyle{every node}=[font=\scriptsize]
  \begin{axis}[
      scatter/classes={
        a={mark=*,blue, scale=0.8},
        b={mark=*,yellow, scale=0.8}},
      width=1\columnwidth,
      height=50mm,
      grid=both,
      legend columns=-1,
      minor x tick num=1,
      minor y tick num=1,
      set layers, 
      mark layer=axis tick labels,
      xlabel=Latency (ns),
      ylabel=Test Error Rate (\%),
      ylabel near ticks,
      tick label style={font=\scriptsize},  
      legend style={at={(0.5,1.2)},anchor=north, font=\scriptsize},
    ]
    \addlegendimage{empty legend}

    \addplot [scatter, only marks, scatter src=explicit symbolic]coordinates {
        (15.912, 4.91) [a]
        (14.259, 5.34) [a] 
        (12.378, 6.07) [a] 
        (10.1, 7.05) [a] 
        (8.868, 8.82) [a] 
        (7.988, 10.88) [a] 
    };

    \addplot [scatter, only marks, scatter src=explicit symbolic] coordinates {
        (15.084, 3.85) [b]
        (12.71, 4.26) [b]
        (10.5, 5.22) [b]
        (8.52, 6.03) [b]
        (7.988, 6.86) [b]
        (7.716, 8.02) [b]
    };

    \node[font=\tiny] at (-10,450) {(80,40,40)};
    \node[font=\tiny] at (2,340) {(120,50,50)};
    \node[font=\tiny] at (80,180) {(200,64,64)};
    \node[font=\tiny] at (270,100) {(256,100,100)};
    \node[font=\tiny] at (500,10) {(256,100,100,100)};
    \node[font=\tiny] at (750,-30) {(256,100,100,100,100)};

    \node[font=\tiny] at (40,740) {(200,64,64)};
    \node[font=\tiny] at (185,530) {(256,100,100)};
    \node[font=\tiny] at (330,345) {(256,100,100,100)};
    \node[font=\tiny] at (580,245) {(256,100,100,100,100)};
    \node[font=\tiny] at (770,170) {(256,150,150,150,150)};
    \node[font=\tiny] at (785,70) {(256,200,200,200,200)};

    \addplot [blue,thick] 
    coordinates {
        (15.912, 4.91)
        (15.912, 5.34)
        (14.259, 5.34)
        (14.259, 6.07)
        (12.378, 6.07)
        (12.378, 7.05)
        (10.1, 7.05)
        (10.1, 8.82)
        (8.868, 8.82)
        (8.868, 10.88)
        (7.988, 10.88)
    };

    \addplot [yellow,thick] 
    coordinates {
        (15.084, 3.85)
        (15.084, 4.26)
        (12.71, 4.26)
        (12.71, 5.22)
        (10.5, 5.22)
        (10.5, 6.03)
        (8.52, 6.03)
        (8.52, 6.86)
        (7.988, 6.86)
        (7.988, 8.02)
        (7.716, 8.02)
    };
    \label{pgfplots:label0}

    \addplot [black, densely dashed, smooth,<-]
    coordinates {
        (10.1, 8.02)
        (8.9, 8.10)
        (7.716, 8.02)
    }
    node[text=black, font=\tiny, above, yshift=-0.1cm] at (1215,28.6) [pos=0.5,font=\scriptsize]{$\mathbf{1.3\times}$};;

    \addplot [black, densely dashed,<-]
    coordinates {
        (12.378, 6.03)
        (10.5, 6.23)
        (8.52, 6.03)
    }
    node[text=black, font=\tiny, below, yshift=0.29cm] at (1215,28.6) [pos=0.5,font=\scriptsize]{$\mathbf{1.5\times}$};;

    \addplot [black, densely dashed,<-]
    coordinates {
        (15.912, 5.22)
        (13.4, 5.32)
        (10.5, 5.22)
    }
    node[text=black, font=\tiny, below, yshift=0.29cm] at (1215,28.6) [pos=0.5,font=\scriptsize]{$\mathbf{1.5\times}$};;

    \addlegendentry{}
    \addlegendentry{\textbf{LogicNets}}
    \addlegendentry{\textbf{NeuraLUT}}
  \end{axis}
\end{tikzpicture}}
	\caption{
            Test error rate vs latency trade-off study, highlighting the top-performing runs across 10 seeds. Each point is labelled with the number of L-LUTs per hidden layer. NeuraLUT features a fixed sub-network of $N$=$16$, $L$=$4$, $S$=$2$.
        }
	\label{fig:par0}
\end{figure}
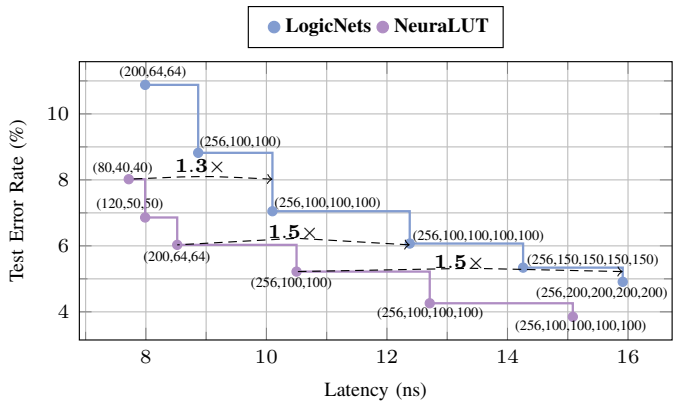
\begin{figure}[tbp]
	\centering
 \hspace{-1.1cm}
        \resizebox{1.1\columnwidth}{!}{\definecolor{yellow}{RGB}{175, 141, 195}
\definecolor{dyellow}{RGB}{136, 52, 171}
\definecolor{blue}{RGB}{130, 156, 208}
\definecolor{red}{RGB}{255, 87, 87}
\definecolor{dblue}{RGB}{82, 113, 255}

\begin{tikzpicture}
  \pgfplotsset{compat=1.5}
  \tikzstyle{every node}=[font=\scriptsize] 
  \begin{axis}[
      scatter/classes={
        a={mark=*,blue, scale=0.8},
        b={mark=*,yellow, scale=0.8}},
      width=1\columnwidth,
      height=50mm,
      grid=both,
      legend columns=-1,
      minor x tick num=1,
      minor y tick num=1,
      set layers, 
      mark layer=axis tick labels,
      xlabel=Area (LUTs),
      ylabel=Test Error Rate (\%),
      tick label style={font=\footnotesize},  
      legend style={at={(0.5,1.2)},anchor=north, font=\footnotesize},
    ]
    \addlegendimage{empty legend}

    \addplot [scatter, only marks, scatter src=explicit symbolic]coordinates {
        (55510, 4.91) [a]
        (42956, 5.34) [a]
        (37133, 6.07) [a]
        (29357, 7.05) [a]
        (24907, 8.82) [a]
        (20226, 10.88) [a]
    };

    \addplot [scatter, only marks, scatter src=explicit symbolic] coordinates {
        (63348, 3.85) [b]
        (52551, 4.26) [b]
        (37974, 5.22) [b]
        (32880, 6.03) [b]
        (23867, 6.86) [b]
        (18450, 8.02) [b]
    };

    \node[font=\tiny] at (-10,450) {(80,40,40)};
    \node[font=\tiny] at (40,270) {(120,50,50)};
    \node[font=\tiny] at (140,180) {(200,64,64)};
    \node[font=\tiny] at (200,100) {(256,100,100)};
    \node[font=\tiny] at (320,10) {(256,100,100,100)};
    \node[font=\tiny] at (430,-40) {(256,100,100,100,100)};

    \node[font=\tiny] at (20,740) {(200,64,64)};
    \node[font=\tiny] at (105,530) {(256,100,100)};
    \node[font=\tiny] at (160,345) {(256,100,100,100)};
    \node[font=\tiny] at (250,245) {(256,100,100,100,100)};
    \node[font=\tiny] at (310,175) {(256,150,150,150,150)};
    \node[font=\tiny] at (410,75) {(256,200,200,200,200)};

    \addplot [blue,thick] 
    coordinates {
        (55510, 4.91)
        (55510, 5.34)
        (42956, 5.34)
        (42956, 6.07)
        (37133, 6.07)
        (37133, 7.05)
        (29357, 7.05)
        (29357, 8.82)
        (24907, 8.82)
        (24907, 10.88)
        (20226, 10.88)
    };

    \addplot [yellow,thick] 
    coordinates {
        (63348, 3.85)
        (63348, 4.26)
        (52551, 4.26)
        (52551, 5.22)
        (37974, 5.22)
        (37974, 6.03)
        (32880, 6.03)
        (32880, 6.86)
        (23867, 6.86)
        (23867, 8.02)
        (18450, 8.02)
    };
    \label{pgfplots:label1}

    \addlegendentry{}
    \addlegendentry{\textbf{LogicNets}}
    \addlegendentry{\textbf{NeuraLUT}}
  \end{axis}
\end{tikzpicture}}
	\caption{
            Test error rate vs area trade-off study for the models in Figure~\ref{fig:par0}. Each point is labelled with the number of L-LUTs per hidden layer. NeuraLUT's improved Pareto frontier highlights that the gains in latency shown in Figure~\ref{fig:par0} are accompanied by a decrease in area cost.
        }
	\label{fig:par1}
\end{figure}
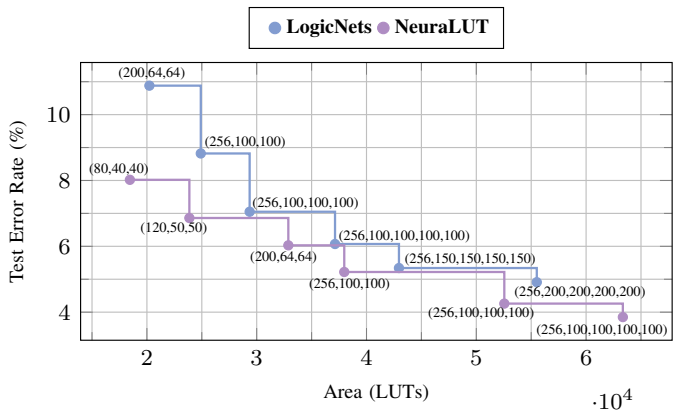

\subsection{MNIST case study}
\subsubsection{Test accuracy}

The primary advantage of training NeuraLUT models lies in their heightened expressibility, resulting in improved test accuracy. To illustrate this advantage, we conduct an analysis on MNIST using a fixed circuit-level model (Figure~\ref{fig:box_plot}). Initially, we train it in the traditional setting hiding single neurons inside the L-LUTs, equivalent to LogicNets. Subsequently, we enhance the model by replacing each neuron with dense sub-networks of increasing depth. Additionally, we evaluate the effectiveness of skip-connections by training NeuraLUT models both with (highlighted in purple) and without them (highlighted in gray).

The key highlight of this analysis is that, for a fixed number of L-LUTs, all NeuraLUT models significantly enhance test accuracy for the same number and size of L-LUTs compared to LogicNets. Furthermore, skip-connections facilitate training to harness the capabilities of deeper sub-networks. Hence, gradually increasing the complexity of the sub-network results in a boost in test accuracy, a benefit that is not seen when skip-connections are omitted. For example, increasing sub-network depth from $L=3$ to $L=4$ leads to a boost in accuracy in the NeuraLUT model, whereas the accuracy drops if the skip-connections are omitted.

\subsubsection{Latency and area}
We have also investigated NeuraLUT's performance in terms of latency and area on the MNIST dataset. Models of different sizes were selected and tested under both LogicNets ($N=1$, $L=1$, $S=0$) and NeuraLUT ($N=16$, $L=4$, $S=2$) settings. Subsequently, the best design points were chosen to construct Pareto frontiers for latency vs. test error rate (Figure~\ref{fig:par0}) and area vs. test error rate (Figure~\ref{fig:par1}). The latency and the area are collected post Place \& Route. It is important to note that each L-LUT layer is evaluated in one clock cycle, therefore the latency of the design is directly proportional to the number of L-LUT layers, which is much lower than the number of neural network layers for NeuraLUT.

In Figure~\ref{fig:par0}, significant latency reductions are evident for NeuraLUT models compared to traditional models for the same test accuracy. Reductions range from $1.3\times$ to up to $1.5\times$. Furthermore, NeuraLUT models demonstrate less pronounced increases in test error rate when the number of circuit-level L-LUTs is reduced. For instance, reducing the network size from $(256,100,100,100,100)$ L-LUTs to $(200,64,64)$ L-LUTs results in a $2.18$ percentage-point increase in error for NeuraLUT compared to a $4.81$ percentage-point increase for traditional models. This highlights the ability of hidden sub-networks in NeuraLUT to mitigate accuracy drops caused by reductions in L-LUTs, suggesting increasing advantages as network width and depth become constrained.

In Figure~\ref{fig:par1}, NeuraLUT's Pareto frontier consistently outperforms the Pareto frontier of the traditional models, highlighting that NeuraLUT is not only latency-efficient, but it is also area-efficient. When the size of the L-LUTs matches or is smaller than that of the P-LUTs on the FPGA, the total LUT utilization for the traditional and NeuraLUT models is the same. However, as the size of the L-LUTs exceeds that of the P-LUTs, the LUT utilization is influenced by the size of the P-LUT circuits generated by the synthesis tools for each L-LUT. We observe in Figure~\ref{fig:par1} that in comparison to traditional implementations, NeuraLUT may require more P-LUTs for the same L-LUT network, as a consequence of the fact that the L-LUTs offer less opportunity for logic simplification when they encode a more complex function. Nevertheless, this increase in LUTs is outweighed by reductions in model size such that for equivalent accuracy, NeuraLUT utilizes less area compared to traditional NNs.

\begin{table*}[tb]
  \caption{Model architectures used for evaluation (Table~\ref{table:evaluation}).}
  \begin{center}
    \renewcommand{\arraystretch}{1.2} 
    \begin{tabular}{ccccccccc}
      \hline
      \textbf{Dataset} & \textbf{Model Name} & \textbf{L-LUTs per Layer} & $\boldsymbol{\beta}$ & \textbf{$F$} & \textbf{$L$} & \textbf{$N$} & \textbf{$S$} & \textbf{Exceptions} \\
      \hline
      \hline
      MNIST & HDR-5L & 256, 100, 100, 100, 10 & 2 & 6 & 4 & 16 & 2 & \\
      \hline
      Jet substructure tagging& JSC-2L & 32, 5 & 4 & 3 & 4 & 8 & 2 & \\
      \hline
      Jet substructure tagging& JSC-5L & 128, 128, 128, 64, 5 & 4 & 3 & 4 & 16 & 2 & $\beta_0=7$, $F_0=2$ \\
      \hline
    \end{tabular}
    \renewcommand{\arraystretch}{1.2} 
    \label{table:networks}
  \end{center}
\end{table*}

\begin{table*}[tb]
  \caption{Evaluation of NeuraLUT on the MNIST and jet substructure tagging datasets. Bold indicates best in class.}
  \begin{center}
    \renewcommand{\arraystretch}{1.5} 
    \begin{tabular}{clrrrrrrrr}
      &&\multirow{2}{*}{\textbf{Accuracy}}&\multirow{2}{*}{\textbf{LUT}}&\multirow{2}{*}{\textbf{FF}}&\multirow{2}{*}{\textbf{DSP}}&\multirow{2}{*}{\textbf{BRAM}}&\textbf{$\text{F}_\text{max}$}&\textbf{Latency}&\textbf{Area$\times$Delay}\\
      &&&&&&&\textbf{(MHz)}&\textbf{(ns)}&\textbf{(LUT}$\times$\textbf{ns)}\\
      \hline
      \cellcolor[gray]{0.9}&\textbf{NeuraLUT (HDR-5L)}&\textbf{96\%}&\textbf{54798}&\textbf{3757}&\textbf{0}&\textbf{0}&\textbf{431}&\cellcolor[gray]{0.9}\textbf{12}&\cellcolor[gray]{0.9}$\mathbf{6.6\times10^5}$\\
      \cline{2-10}
      \cellcolor[gray]{0.9}&\textbf{PolyLUT}~\cite{poly}&\textbf{96\%}&70673&4681&\textbf{0}&\textbf{0}&378&\cellcolor[gray]{0.9}16&\cellcolor[gray]{0.9}$11.3\times10^5$\\
      \cline{2-10}
      \cellcolor[gray]{0.9}&\textbf{FINN}~\cite{finn}&\textbf{96\%}&91131&-&\textbf{0}&5&200&\cellcolor[gray]{0.9}310&\cellcolor[gray]{0.9}$282.5\times10^5$\\
      \cline{2-10}
      \multirow{-4}{*}{\cellcolor[gray]{.9}\textbf{MNIST}} &\textbf{\texttt{hls4ml}}~\cite{hls4ml}&95\%&260092&165513&\textbf{0}&\textbf{0}&200&\cellcolor[gray]{0.9}190&\cellcolor[gray]{0.9}$494.2\times10^5$\\
      \hline\hline
      \cellcolor[gray]{0.9}&\textbf{NeuraLUT (JSC-2L)}&\textbf{72\%}&\textbf{4684}&\textbf{341}&\textbf{0}&\textbf{0}&\textbf{727}&\cellcolor[gray]{0.9}\textbf{3}&\cellcolor[gray]{0.9}$\mathbf{1.4\times10^4}$\\
      \cline{2-10}
      \multirow{0.8}{*}{\cellcolor[gray]{.9}\textbf{Jet substructure tagging}}&\textbf{PolyLUT}~\cite{poly}&\textbf{72\%}&12436&773&\textbf{0}&\textbf{0}&646&\cellcolor[gray]{0.9}5&\cellcolor[gray]{0.9}$6.2\times10^4$\\
      \cline{2-10}
      \multirow{0.8}{*}{\cellcolor[gray]{.9}\textbf{(low accuracy)}}&\textbf{LogicNets}~\cite{logicnets}$^{\mathrm{a}}$&\textbf{72\%}&37931&810&\textbf{0}&\textbf{0}&427&\cellcolor[gray]{0.9}13&\cellcolor[gray]{0.9}$49.3\times10^4$\\
      \hline\hline
      \cellcolor[gray]{0.9}&\textbf{NeuraLUT (JSC-5L)}&75\%&92357&4885&\textbf{0}&\textbf{0}&\textbf{368}&\cellcolor[gray]{0.9}\textbf{14}&\cellcolor[gray]{0.9}$\mathbf{1.3\times10^6}$\\
      \cline{2-10}
      \cellcolor[gray]{0.9}&\textbf{PolyLUT}~\cite{poly}&75\%&236541&2775&\textbf{0}&\textbf{0}&235&\cellcolor[gray]{0.9}21&\cellcolor[gray]{0.9}$5\times10^6$\\
      \cline{2-10}
       \multirow{-3}{*}{\cellcolor[gray]{.9}\textbf{Jet substructure tagging}}&\textbf{Duarte \textit{et al.}}~\cite{duarte}&75\%&\multicolumn{2}{c}{88797$^{\mathrm{b}}$}&954&\textbf{0}&200&\cellcolor[gray]{0.9}75&\cellcolor[gray]{0.9}$6.7\times10^6$\\
      \cline{2-10}
      \multirow{-3}{*}{\cellcolor[gray]{.9}\textbf{(high accuracy)}} &\textbf{Fahim \textit{et al.}}~\cite{fahim}&\textbf{76\%}&\textbf{63251}&\textbf{4394}&38&\textbf{0}&200&\cellcolor[gray]{0.9}45&\cellcolor[gray]{0.9}$2.8\times10^6$\\
      \hline
      \multicolumn{9}{l}{$^{\mathrm{a}}$New results can be found on the LogicNets GitHub page.}\\
      \multicolumn{9}{l}{$^{\mathrm{b}}$Paper reports ``LUT+FF".}\\
    \end{tabular}
    \renewcommand{\arraystretch}{1.5}
    \label{table:evaluation}
  \end{center}
  \vspace{-0.1cm}
\end{table*}

\subsection{Comparison with prior work}
We assess NeuraLUT based on accuracy, logic utilization, maximum frequency, and latency. However, our methodology targets edge applications with a shared emphasis on low latency and area, hence we also evaluate by area-delay product.

We trained NeuraLUT using various model architectures and, for fair comparison, we used those NeuraLUT parameters giving rise to comparable test accuracy to prior works, PolyLUT~\cite{poly}, LogicNets~\cite{logicnets}, FINN~\cite{finn}, \texttt{hls4ml}~\cite{hls4ml}, Duarte \textit{et al.}~\cite{duarte}, and Fahim \textit{et al.}~\cite{fahim}, aiming to optimize latency and area utilization. These models are detailed in Table~\ref{table:networks}. NullaNet~\cite{nullanet} cannot be a direct point of comparison since it implements the first and last layers in floating-point and reports hardware results only for the hidden layers. Additionally, LogicNets does not benchmark on MNIST.
\subsubsection{MNIST}
In our evaluation, we benchmarked NeuraLUT's performance on the MNIST dataset against state-of-the-art results from existing work on ultra-low latency implementations. We compared it with the performance of PolyLUT on the HDR model~\cite{poly}, the values reported by FINN~\cite{finn} on the SFC-max model (a binary and fully unfolded implementation), and the results of \texttt{hls4ml}~\cite{duarte} utilizing a ternary neural network model. The evaluation outcomes are detailed in Table~\ref{table:evaluation}.

For this dataset, we employed the HDR-5L model, featuring a $5$-layer circuit-level architecture, comprising $4$-layer sub-networks with $2$ skip-connections. Achieving the same accuracy or more, we outperform all prior work across all metrics. Compared to PolyLUT, FINN, and \texttt{hls4ml}, we achieved significant reductions in the area-delay product by $1.7\times$, $42.8\times$, and $74.9\times$, respectively. Our advantage over FINN and \texttt{hls4ml} is attributed to concealing all the computational components and dense parts of the network within LUTs, therefore reducing the number of logic and exposed datapaths to a minimum. Against PolyLUT, we achieve improved implementations due to the efficient handling of function complexity within the L-LUTs.

\subsubsection{Jet substructure tagging}
For the evaluation of NeuraLUT on the jet substructure tagging dataset, we divided our analyses into two segments aimed at different test accuracies (Table~\ref{table:evaluation}). The first segment focuses on achieving a lower test accuracy, comparing our method with PolyLUT's JSC-M Lite model~\cite{poly} and LogicNets' JSC-M model~\cite{logicnets}. In the second segment, we evaluate our approach against PolyLUT's HDR model~\cite{poly}, as well as implementations proposed by Duarte \textit{et al.}~\cite{duarte} and Fahim \textit{et al.}~\cite{fahim}.

NeuraLUT's JSC-2L, comprising only two very shallow layers, achieves the same accuracy as PolyLUT and LogicNets while achieving significant reductions of $4.4\times$ and $35.2\times$ in the area-delay product, respectively. This aligns with the findings of the case study, underscoring the efficacy of NeuraLUT's highly expressive L-LUTs in restoring precision within highly constrained circuit-level networks.

NeuraLUT's JSC-5L reaches the lowest latency and is more efficient in minimizing the area-delay product compared to PolyLUT, the work by Duarte \textit{et al.}~\cite{duarte} and the work of Fahim \textit{et al.}~\cite{fahim} reaching reductions of $3.8\times$, $5.2\times$, and $2.2\times$, respectively. NeuraLUT achieves this without utilizing DSPs at the expense of slightly higher LUT utilization compared to the works of Duarte \textit{et al.}~\cite{duarte} and Fahim \textit{et al.}~\cite{fahim}.

In summary, on all benchmarks, NeuraLUT archives the smallest area-delay product and reports latency reductions of up to $26\times$ on MNIST and $5\times$ on the jet substructure tagging dataset.

\section{Conclusion and Further Work}
Our work introduces a novel approach to LUT-based DNN acceleration, which involves mapping entire sub-networks of arbitrary topology to L-LUTs rather than individual neurons as seen in prior works. This strategy offers greater accuracy to LUT-based models, resulting in lower-latency networks with improved function expressivity. By incorporating skip-connections within partitions, we mitigate challenges like vanishing gradients, enabling the training of deeper sub-networks while maintaining efficiency. Our proposed methodology is validated through experiments on latency-critical tasks such as jet substructure tagging and digit classification using MNIST, showcasing significant improvements in the area-delay product while preserving accuracy.

Although NeuraLUT's sub-networks within a partition are dense and full precision, they are limited to a small number of low precision inputs and outputs at partition boundaries. This constraint originates from the exponential scaling of L-LUTs, inherited from traditional LUT-based approaches. Consequently, restrictions remain on both the number of inputs and precision, making NeuraLUT mainly suitable, like LogicNets and PolyLUT, for small ultra-low latency embedded applications.

A possible next step is to explore automated search techniques like Neural Architecture Search (NAS) to optimize NeuraLUT's circuit-level topology or sub-network topology. NAS can maximize performance while addressing L-LUT constraints, enhancing efficiency, and unlocking the full potential of NeuraLUT's flexibility.

\clearpage
\bibliographystyle{IEEEtran}
\begingroup
\raggedright
\bibliography{bibs}

\begin{thebibliography}{10}
\providecommand{\url}[1]{#1}
\csname url@samestyle\endcsname
\providecommand{\newblock}{\relax}
\providecommand{\bibinfo}[2]{#2}
\providecommand{\BIBentrySTDinterwordspacing}{\spaceskip=0pt\relax}
\providecommand{\BIBentryALTinterwordstretchfactor}{4}
\providecommand{\BIBentryALTinterwordspacing}{\spaceskip=\fontdimen2\font plus
\BIBentryALTinterwordstretchfactor\fontdimen3\font minus \fontdimen4\font\relax}
\providecommand{\BIBforeignlanguage}[2]{{%
\expandafter\ifx\csname l@#1\endcsname\relax
\typeout{** WARNING: IEEEtran.bst: No hyphenation pattern has been}%
\typeout{** loaded for the language `#1'. Using the pattern for}%
\typeout{** the default language instead.}%
\else
\language=\csname l@#1\endcsname
\fi
#2}}
\providecommand{\BIBdecl}{\relax}
\BIBdecl

\bibitem{duarte}
{J. Duarte \emph{et al.}}, ``{Fast inference of deep neural networks in FPGAs for particle physics},'' \emph{{Journal of Instrumentation}}, vol.~13, no.~7, p. P07027, Jul. 2018, doi: 10.1088/1748-0221/13/07/P07027.

\bibitem{murovic}
T.~Murovič and A.~Trost, ``{Genetically optimized massively parallel binary neural networks for intrusion detection systems},'' \emph{{Computer Communications}}, vol. 179, no.~7, pp. 1--10, Nov. 2021, doi: 10.1016/j.comcom.2021.07.015.

\bibitem{cookiebox}
A.~Corbeil~Therrien, R.~Herbst, O.~Quijano, A.~Gatton, and R.~Coffee, ``{Machine Learning at the Edge for Ultra High Rate Detectors},'' in \emph{{2019 IEEE Nuclear Science Symposium and Medical Imaging Conference (NSS/MIC)}}.\hskip 1em plus 0.5em minus 0.4em\relax IEEE, Oct. 2019, pp. 1--4, doi: 10.1109/NSS/MIC42101.2019.9059671.

\bibitem{edge}
M.~M.~H. Shuvo, S.~K. Islam, J.~Cheng, and B.~I. Morshed, ``{Efficient Acceleration of Deep Learning Inference on Resource-Constrained Edge Devices: A Review},'' \emph{{Proceedings of the IEEE}}, vol. 111, no.~1, pp. 42--91, Jan. 2023, doi: 10.1109/JPROC.2022.3226481.

\bibitem{esurvey}
{E. Wang \emph{et al.}}, ``{Deep Neural Network Approximation for Custom Hardware},'' \emph{{ACM Computing Surveys}}, vol.~52, no.~2, pp. 1--39, May 2019, doi: 10.1145/3309551.

\bibitem{lutnet1}
E.~Wang, J.~J. Davis, P.~Y.~K. Cheung, and G.~A. Constantinides, ``{LUTNet: Rethinking Inference in FPGA Soft Logic},'' in \emph{{2019 IEEE 27th Annual International Symposium on Field-Programmable Custom Computing Machines (FCCM)}}.\hskip 1em plus 0.5em minus 0.4em\relax {San Diego, CA, USA}: {IEEE Computer Society}, May 2019, pp. 26--34, doi: 10.1109/FCCM.2019.00014.

\bibitem{poly}
M.~Andronic and G.~A. Constantinides, ``{PolyLUT: Learning Piecewise Polynomials for Ultra-Low Latency FPGA LUT-based Inference},'' in \emph{{2023 International Conference on Field Programmable Technology (ICFPT)}}.\hskip 1em plus 0.5em minus 0.4em\relax Yokohama, Japan: IEEE, Dec. 2023, pp. 60--68, doi: 10.1109/ICFPT59805.2023.00012.

\bibitem{logicnets}
Y.~Umuroglu, Y.~Akhauri, N.~J. Fraser, and M.~Blott, ``{LogicNets: Co-Designed Neural Networks and Circuits for Extreme-Throughput Applications},'' in \emph{{2020 30th International Conference on Field-Programmable Logic and Applications (FPL)}}.\hskip 1em plus 0.5em minus 0.4em\relax {Gothenburg, Sweden}: {IEEE Computer Society}, Sep. 2020, pp. 291--297, doi: 10.1109/FPL50879.2020.00055.

\bibitem{nullanet}
M.~Nazemi, G.~Pasandi, and M.~Pedram, ``{Energy-Efficient, Low-Latency Realization of Neural Networks through Boolean Logic Minimization},'' in \emph{{Proceedings of the 24th Asia and South Pacific Design Automation Conference}}.\hskip 1em plus 0.5em minus 0.4em\relax {Tokyo, Japan}: Association for Computing Machinery, Jan. 2019, p. 274–279, doi: 10.1145/3287624.3287722.

\bibitem{fahim}
\BIBentryALTinterwordspacing
{F. Fahim \emph{et al.}}, ``{hls4ml: An Open-Source Codesign Workflow to Empower Scientific Low-Power Machine Learning Devices},'' in \emph{{TinyML Research Symposium’21}}, {San Jose, CA, USA}, Mar. 2021. [Online]. Available: \url{https://doi.org/10.48550/arXiv.2103.05579}
\BIBentrySTDinterwordspacing

\bibitem{summers}
{S. Summers \emph{et al.}}, ``{Fast inference of Boosted Decision Trees in FPGAs for particle physics},'' \emph{{Journal of Instrumentation}}, vol.~15, no.~5, p. P05026, May 2020, doi: 10.1088/1748-0221/15/05/P05026.

\bibitem{coelho}
{C. N. Coelho, Jr. \emph{et al.}}, ``{Automatic heterogeneous quantization of deep neural networks for low-latency inference on the edge for particle detectors},'' \emph{{Nature Machine Intelligence}}, vol.~3, no.~6, pp. 675--686, Jun. 2021, doi: 10.1038/s42256-021-00356-5.

\bibitem{finn}
{Y. Umuroglu \emph{et al.}}, ``{FINN: A Framework for Fast, Scalable Binarized Neural Network Inference},'' in \emph{{Proceedings of the 2017 ACM/SIGDA International Symposium on Field-Programmable Gate Arrays}}, ser. {FPGA '17}.\hskip 1em plus 0.5em minus 0.4em\relax {Monterey, CA, USA}: {Association for Computing Machinery}, Feb. 2017, p. 65–74, doi: 10.1145/3020078.3021744.

\bibitem{hls4ml}
{J. Ngadiuba \emph{et al.}}, ``{Compressing deep neural networks on FPGAs to binary and ternary precision with hls4ml},'' \emph{{Machine Learning: Science and Technology}}, vol.~2, no.~1, p. 015001, Dec. 2020, doi: 10.1088/2632-2153/aba042.

\bibitem{lutnet2}
E.~Wang, J.~J. Davis, P.~Y.~K. Cheung, and G.~A. Constantinides, ``{LUTNet: Learning FPGA Configurations for Highly Efficient Neural Network Inference},'' \emph{{IEEE Transactions on Computers}}, vol.~69, no.~12, pp. 1795--1808, Dec. 2020, doi: 10.1109/TC.2020.2978817.

\bibitem{expander}
A.~Prabhu, G.~Varma, and A.~Namboodiri, ``{Deep Expander Networks: Efficient Deep Networks from Graph Theory},'' in \emph{{Computer Vision - ECCV 2018}}, V.~Ferrari, M.~Hebert, C.~Sminchisescu, and Y.~Weiss, Eds.\hskip 1em plus 0.5em minus 0.4em\relax {Munich, Germany}: {Springer International Publishing}, Oct. 2018, pp. 20--36, doi: 10.1007/978-3-030-01261-8\_2.

\bibitem{NIN}
M.~Lin, Q.~Chen, and S.~Yan, ``{Network In Network},'' in \emph{{International Conference on Learning Representations (ICLR) 2014}}, {Banff, AB, Canada}, Apr. 2014, pp. 1--10.

\bibitem{cybenko}
G.~V. Cybenko, ``{Approximation by Superpositions of a Sigmoidal Function},'' \emph{{Mathematics of Control, Signals and Systems}}, vol.~2, pp. 303--314, Dec. 1989.

\bibitem{hornik}
K.~Hornik, ``{Approximation Capabilities of Muitilayer Feedforward Networks},'' \emph{{NEURAL NETWORKS}}, vol.~4, no.~2, pp. 251--257, 1991.

\bibitem{bengio}
Y.~Bengio, P.~Simard, and P.~Frasconi, ``{Learning long-term dependencies with gradient descent is difficult},'' \emph{{IEEE Transactions on Neural Networks}}, vol.~5, no.~2, pp. 157--66, Mar. 1994, doi: 10.1109/72.279181.

\bibitem{glorot}
X.~Glorot and Y.~Bengio, ``{Understanding the difficulty of training deep feedforward neural networks},'' in \emph{{Proceedings of the Thirteenth International Conference on Artificial Intelligence and Statistics}}, ser. {Proceedings of Machine Learning Research}, Y.~W. Teh and M.~Titterington, Eds., vol.~9.\hskip 1em plus 0.5em minus 0.4em\relax Sardinia, Italy: PMLR, May 2010, pp. 249--256.

\bibitem{resnet}
K.~He, X.~Zhang, S.~Ren, and J.~Sun, ``{Deep residual learning for image recognition},'' in \emph{{2016 IEEE Conference on Computer Vision and Pattern Recognition (CVPR)}}, {Los Alamitos, CA, USA}, Jun. 2016, pp. 770--778, doi 10.1109/CVPR.2016.90.

\bibitem{reg}
\BIBentryALTinterwordspacing
I.~Loshchilov and F.~Hutter, ``{Fixing Weight Decay Regularization in Adam},'' in \emph{{7th International Conference on Learning Representations}}, {New Orleans, LA, USA}, May 2019. [Online]. Available: \url{https://openreview.net/forum?id=Bkg6RiCqY7}
\BIBentrySTDinterwordspacing

\bibitem{sgd}
\BIBentryALTinterwordspacing
------, ``{SGDR: Stochastic Gradient Descent with Warm Restarts},'' in \emph{{5th International Conference on Learning Representations}}, {Toulon, France}, Apr. 2017. [Online]. Available: \url{https://openreview.net/forum?id=Skq89Scxx}
\BIBentrySTDinterwordspacing

\bibitem{brevitas}
\BIBentryALTinterwordspacing
A.~Pappalardo, ``{Brevitas: neural network quantization in PyTorch},'' Jun. 2020, {[Accessed: January 12, 2024]}. [Online]. Available: \url{https://github.com/Xilinx/brevitas}
\BIBentrySTDinterwordspacing

\bibitem{mnist}
\BIBentryALTinterwordspacing
Y.~LeCun, C.~Cortes, and C.~J. Burges, ``{The MNIST database of handwritten digits}.'' [Online]. Available: \url{http://yann.lecun.com/exdb/mnist/}
\BIBentrySTDinterwordspacing

\end{thebibliography}
\endgroup

\end{document}